\documentclass{IEEEtran}

\usepackage{graphicx}
\usepackage{amsmath}
\usepackage{amssymb}

\newcommand{\h}{\frac{1}{2}}
\newcommand{\s}[1]{\ensuremath{\mathcal{#1}}}
\newcommand{\f}[1]{\ensuremath{ \lfloor #1 \rfloor }}
\renewcommand{\QED}{\hfill $\Box$}
\newtheorem{Theorem}{Theorem}
\newtheorem{Lemma}{Lemma}
\newenvironment{Proof}[1]{\noindent \textbf{Proof#1: }}{\QED}
\begin{document}

\newpage \onecolumn
This manuscript is the accepted version of the following paper published by the IEEE:\\

"Asymptotic Capacity Bounds for Wireless Networks with Non-Uniform Traffic Patterns," S. Toumpis, IEEE Trans. Wireless Comm., vol. 7, no. 6, pp. 2231-2242, June 2008.\\

DOI: 10.1109/TWC.2008.061010  

IEEE Xplore link: http://ieeexplore.ieee.org/xpl/articleDetails.jsp?arnumber=4543075 \\

The following copyright notice applies:\\
" \copyright 2008 IEEE. Personal use of this material is permitted. Permission from IEEE must be obtained for all other uses, in any current or future media, including reprinting/republishing this material for advertising or promotional purposes, creating new collective works, for resale or redistribution to servers or lists, or reuse of any copyrighted component of this work in other works."

\twocolumn
\newpage

\author{Stavros Toumpis, \IEEEmembership{Member, IEEE}}
\title{Asymptotic Capacity Bounds for Wireless \\ Networks with Non-Uniform Traffic Patterns \thanks{ \noindent The author is with the Department of Electrical and Computer Engineering of the University of Cyprus, Kallipoleos 75, P.O. Box 20537, 1678 Nicosia, Cyprus. Part of this work was conducted while the author was at the Telecommunications Research Center Vienna (ftw.), in Vienna, Austria. Work was supported by \textbf{K}\emph{plus} funding for the ftw. project I0 ``Signal and Information Processing'' and FP6 IST funding through the NetReFound project. A preliminary version of Sections \ref{section:asymmetric}, \ref{section:cluster}, and \ref{section:hybrid} appeared in \cite{toumpis13}. The basic result of Section~\ref{section:asymmetric} also appeared, with sketches of proofs, in \cite{toumpis12}. The basic result of Section \ref{section:multicast} appeared, with a sketch of the proof, in the oral presentation of \cite{toumpis13}.}}
\maketitle

\vspace{-0.7in}
\begin{abstract} We develop bounds on the capacity of wireless multihop networks when the traffic pattern is non-uniform, i.e., not all nodes are the sources and sinks of similar volumes of traffic. Our results are asymptotic, i.e., they hold with probability going to unity as the number of nodes goes to infinity. We study \emph{(i)} asymmetric networks, where the numbers of sources and destinations of traffic are unequal, \emph{(ii)} multicast networks, in which each created packet has multiple destinations, \emph{(iii)} cluster networks, that consist of clients and a limited number of cluster heads, and each client wants to communicate with any one of the cluster heads, and \emph{(iv)} hybrid networks, in which the nodes are supported by a limited infrastructure. Our findings quantify the fundamental capabilities of these wireless multihop networks to handle traffic bottlenecks, and point to correct design principles that achieve the capacity without resorting to overly complicated protocols.

\end{abstract}
\thispagestyle{empty}

\section{Introduction}
\label{section:introduction}

We study the setting in which nodes equipped with wireless transceivers communicate over a shared wireless channel to create a multihop network. In this context, we develop bounds on the capacity of the network, which is defined as the theoretical limit on the total traffic that the network can support, assuming optimal coordination among the nodes. The bounds are determined assuming a number of different non-uniform traffic pattern models under which some nodes are required to either create or receive much more traffic than other nodes. Following the approach introduced in~\cite{gupta1}, our results are asymptotic, i.e., they hold with probability going to unity as the number of nodes goes to infinity. 

In~\cite{gupta1}, the authors consider a set of $n$ nodes randomly placed on the surface of a sphere. Each of the nodes chooses another node as the destination for its traffic, randomly, uniformly and independently, and all $n$ traffic streams are assumed to have a common rate requirement. The authors aim to find the maximum possible rate per stream $\lambda(n)$ that the network can achieve. Note that, because the placement of the nodes and the choice of destinations are random, $\lambda(n)$ is a random variable. The authors show that \textbf{with high probability (w. h. p.)}, i.e., with probability going to $1$ as the number of nodes $n$ goes to infinity, $ \frac{K_1}{\sqrt{n \log n}} < \lambda(n) < \frac{K_2}{\sqrt{n \log n}}$, for some $K_2>K_1>0$. Therefore, w. h. p., the maximum possible aggregate throughput $n \lambda(n)$ is \textbf{on the order of} the square root of the nodes $\sqrt{n}$, i.e., ignoring poly-logarithmic factors of the form $k_1 (\log n)^{k_2}$, the aggregate throughput increases with $n$ as $\sqrt{n}$. As a by-product of our contributions, we offer in the appendix a simple proof for the lower bound, in a setting similar to that of~\cite{gupta1}. Many researchers have followed the same tangent, and a significant number of results of the same flavor have accumulated~\cite{negi1, peraki1, kozat1, towsley1, li1, perevalov3, perevalov6, jacquet2, neely2, sharma1, toumpis9}. 

The traffic pattern used in \cite{gupta1} and almost all of the following works is, in some sense, as simple as possible: All nodes create data with the same rate $\lambda(n)$, and each of them picks at random one of the rest of the nodes as the destination for these data. For lack of a better description, we call this traffic pattern \textbf{uniform}. The uniform traffic pattern is a good model for certain networks, for example those used to support unicast voice transmission. On the other hand, there is a host of applications in which the traffic patterns will be fundamentally different. For example, in a network designed to support multimedia traffic between soldiers in a battlefield most of the traffic will have multiple destinations. As another example, in typical wireless sensor networks, a large number of sensors is interested in communicating with a relatively small number of sinks. However, the asymptotic properties of the capacity under such non-uniform traffic patterns remain to a large extend unexplored. A few notable exceptions are \cite{kozat1, towsley1, li1, perevalov6, jacquet2}. (Of these works, \cite{towsley1} considers localized traffic and \cite{perevalov6} considers a topology with only two clusters, and so their results are unrelated to our results. The relation of our work to each of the rest is clarified later in the text.)

\subsection{Contributions}

In this work, we study wireless networks with no less than four different non-uniform traffic patterns, which collectively cover a wide variety of scenarios. In the process, we develop a versatile methodology that can easily be extended to other traffic patterns as well, such as the traffic pattern with localized traffic of~\cite{towsley1}, which we do not study here. We calculate bounds on the capacity, which we define as the maximum possible aggregate throughput that the network can support under an optimal coordination of the nodes. Following the approach of \cite{gupta1}, our results are asymptotic, i.e., they only hold with probability going to $1$ as the number of nodes $n$ goes to infinity. In addition, we bound the capacity only up to the exponent of $n$, where $n$ is the number of nodes in the networks. In the interest of brevity, we focus on constructive lower bounds, and formally derive only a few upper bounds. In addition, we also present a few other non-critical upper bounds with no formal proofs, but with strong heuristic justifications.

Note that, although we are inspired by the results of~\cite{gupta1}, we do not use them. Moreover, in contrast to \cite{gupta1}, we achieve our results using basic tools of probability and a simple methodology, introduced in~\cite{toumpis9} and extended here, which determines the rate with which the probability converges to unity. We also use a realistic channel model that includes a general model for flat fading. 

We first study \textbf{asymmetric networks}. These consist of two types of nodes: $n$ source nodes, and $n^d $ destination nodes\footnote{Note that formally $n^d$ must be an integer, which only occurs for certain combinations of $n$ and $d$. However in the following we will ignore this and similar issues, as a more formal treatment, for example by using $\lfloor n^d \rfloor$, i.e., the integer part of $n^d$, would encumber the notation without affecting the essence of the derivations.}, where $0 < d < 1$. Sources create packets with a common data rate, and the packets of each source must be delivered to a single one of the destinations, chosen at random. Our main find is that when $d < \h$, there are so few destinations, that bottlenecks start to form around them, constraining the maximum possible aggregate throughput to be around $n^d$. If, on the other hand, $d>\h$, bottlenecks can be avoided, and the capacity is on the order of $n^\h$, as in the uniform traffic setting of \cite{gupta1}.

We then consider \textbf{multicast networks}. These consist of $n$ nodes, each creating packets with a common data rate. Each packet must be delivered to $n^d$ distinct nodes chosen randomly among the rest. (Again, $0<d<1$.) In this context, the capacity is on the order of $n^\frac{d+1}{2}$, and can be achieved without any multicasting in the media access layer, and using a multicast routing tree that can be constructed using only local information.

We also study \textbf{cluster networks}, which consist of $n$ cluster nodes and $n^d$ cluster heads, where $0<d<1$. Each cluster node is the source of a traffic stream and the sink of a traffic stream. The traffic must be between the node and \emph{any} of the cluster heads, and all traffic streams have a common data rate. We show that the maximum possible aggregate throughput is on the order of $n^d$, and can be achieved (up to the order) without routing along multiple hops, even in the presence of fading.

We conclude by studying \textbf{hybrid networks}, continuing the work of~\cite{kozat1, towsley1}. These consist of $n$ wireless nodes and $n^d$ access points, where $0<d<1$. Access points are equipped with wireless transceivers that are identical to the transceivers carried by the wireless nodes. In addition, they are connected with each other through an independent network of practically infinite capacity. Each of the wireless nodes is creating traffic destined for one of the other wireless nodes, chosen at random. The access points have no traffic requirements of their own, but are there to support the communication of the wireless nodes. In this setting, we find that if $d<\h$, then there are so few access points that if the wireless nodes attempt to use them to route a substantial part of their traffic, bottlenecks will be created. Therefore, it is better for the wireless nodes to simply ignore the presence of the access points, and communicate with each other exclusively over the wireless channel. It follows that the capacity is on the order of $n^\frac{1}{2}$. If, however, $d>\h$, there is a sufficient number of access points to make a difference, and the capacity is on the order of $n^d$. Furthermore this capacity can be achieved without multihop wireless communication between wireless nodes.

The rest of this paper is organized as follows: in Section~\ref{section:model} we specify our network models and formally present our results. In Section~\ref{section:lemmas} we present three lemmas that will be used throughout the text. Proofs for the results for asymmetric, multicast, cluster, and hybrid networks are developed in Sections~\ref{section:asymmetric}, \ref{section:multicast}, \ref{section:cluster}, and \ref{section:hybrid} respectively. We conclude in Section~\ref{section:conclusions}. In the appendix, we have included a proof for the lower bound of~\cite{gupta1} we mentioned, in a setting similar to that of~\cite{gupta1}. The proof is included here for reasons of completeness, and, as it uses parts of the proofs of the other results, it is very short.

\section{Network Models and Results}
\label{section:model}

\subsection{Channel and Physical Layer Models}
\label{section:channel}
Nodes are equipped with transceivers used for communication over a wireless channel of bandwidth $W$, and cannot transmit and receive simultaneously. Each node $Z_i$ can transmit with any power $P_i \leq P_0$, where $P_0$ is a global maximum. When $Z_i$ transmits with power $P_i$, $Z_j$ receives the transmitted signal with power $G_{ij} P_i$, where $G_{ij}=K f_{ij} |Z_i-Z_j|^{-\alpha}$. $K$ is a constant, the same for all nodes, $|Z_i-Z_j|$ is the distance between nodes $Z_i$ and $Z_j$, $\alpha > 2$ is the \textbf{decay exponent}, and the factor $f_{ij}$ is the \textbf{fading coefficient}, a non-negative random variable that models fading. 

We assume that the expectation $E[f_{ij}]=1$, and that $f_{ij}=f_{ji}$. Distinct fading coefficients are independent and identically distributed (iid). We also assume that:
\begin{equation}
F^c(x) \triangleq P[f_{ij}>x] \leq \exp[-q x]~\forall x > x_1, \label{eq1}
\end{equation}
for some $q,~x_1>0$. In other words, the complementary cumulative distribution function of the fading distribution has an exponentially thin tail. Intuitively speaking, very high values for the fading coefficients are very rare. Also, we assume that there is a median value $f_M>0$ such that $P[f_{ij} \geq f_M] \geq \h$. Both of these assumptions are satisfied by most distributions used to model fading, for example the Nakagami, Ricean, and Rayleigh distributions, and the trivial distribution for which $P[f_{ij}=1]=1$.

Let $\{Z_t: t \in \s{T} \}$ be the transmitting nodes at a given time, node $Z_t$ transmitting with power $P_t$. Let us assume that node $Z_j$, $j \not \in \s{T}$, is receiving a data packet from $Z_i$, $i \in \s{T}$. Then the \textbf{Signal to Interference and Noise Ratio (SINR)} at node $Z_j$ will be $\gamma_j=\frac{G_{i j}P_i}{\eta + \sum_{k \in \s{T},~k \neq i} G_{k j}P_k}$, where $\eta$ is the receiver thermal noise power, same for all nodes. The transmission will be successful if and only if, for the whole period of transmission, the transmission rate used, $R_j$, satisfies the inequality 
\begin{equation*}
R_j \leq f_R(\gamma_j) \triangleq W \log_2(1+\frac{1}{\Gamma} \gamma_j).
\end{equation*} 
For various values of $\Gamma>1$,  the equation approximates the maximum rate that meets a given BER requirement under a variety of modulation and coding schemes~\cite{forney1}. With $\Gamma=1$, it gives the Shannon bound. We can think of  $f_R(\gamma_j)$ as a function modeling the capabilities of the receiver.

We do not make any additional assumption regarding the physical layer. We note, however, that we implicitly assumed a single common wireless channel. For lack of space we do not consider the case of having multiple channels, but it is intuitively clear that, had there been multiple channels, our capacity calculations would hold in each of them, and the capacity would not change. (See \cite{gupta1} for a formal development of this argument.) Also, under our current model, nodes try to decode only the signal of one transmitter, therefore cooperative communication schemes are excluded. Such schemes have recently been shown to have dramatic capacity improvements~\cite{ozgur1}. However in our work we focus on the effects of traffic asymmetries, and for this reason we keep the physical layer relatively simple. 

We also place no particular restriction on the access scheme used by the nodes. The nodes are free to use, for example, random access schemes such as Aloha, Code Division Multiple Access (CDMA), Time Division Multiple Access (TDMA), or any other access scheme they choose. However, in our constructive lower bounds, we use simple TDMA schemes that, as we show, operate very close to the capacity.

\subsection{Asymmetric Networks}

\textbf{Asymmetric networks} consist of $n$ \textbf{source nodes} $X_1$, $X_2$, $\ldots$, $X_n$, and $m(n)=n^d$ \textbf{destination nodes} $Y_1$, $Y_2$, $\ldots$, $Y_m$, placed randomly, and in particular uniformly and independently, in the unit square $\{(x,y):  |x|,|y| \leq \h\}$. We call $d \in (0,1)$ the \textbf{destination exponent}. Each source node is creating data traffic with a fixed data rate $\lambda(n) ~\mathrm{bps}$, common for all sources, that must be delivered to one of the destination nodes. Each source selects its destination randomly, again uniformly and independently of the others. Both types of nodes are allowed to transmit, receive, and relay packets.

The fundamental difference of this network from previously considered networks, such as the one in~\cite{gupta1}, is not that there are two types of nodes (sources and destinations), but the fact that their numbers $n$ and $m(n)$ are different, and so the traffic pattern is asymmetric: on the average more packets must arrive at each destination, than there are leaving each source. In fact, as will become intuitively clear, we could have assumed, just as well, that there are $n$ destination nodes and only $m(n)=n^d$ source nodes, and arrived at essentially the same results. Applications where traffic pattern asymmetries are expected are, for example, vehicular ad hoc networks in which many users will be downloading infotainment from a few central locations, and wireless sensor networks where the sensor nodes will be exchanging data with a small number of sinks. 

We define the \textbf{capacity $C(n)$ of the network} as the supremum of all rates $\lambda(n)$ that are uniformly achievable by all sources, multiplied by their number $n$. Since the locations of the nodes, the destination of each data stream, and the fading coefficients are random, the capacity is a random variable.

\begin{Theorem}
\emph{In asymmetric networks the capacity $C(n)$ is bounded with high probability (w. h. p.), i.e., with probability approaching unity as $n$ goes to infinity, as follows:
\begin{eqnarray}
C(n) &\leq& \left[ \frac{4 \alpha W}{\log 2} \right] n^d \log n, \label{eq2a} \\
C(n) &\geq& D \times \begin{cases} \frac{\sqrt{2}}{27} \frac{n^\h} {(\log n)^\frac{3}{2}} & \mbox{if  }~ \h < d < 1, \vspace{0.1in} \\
\left[ \frac{1-2d}{5}\right] \frac{n^d}{ \log n} & \mbox{if  }~ 0 < d <\h, \end{cases}  \label{eq2b}
\end{eqnarray}
where the constant $D$ is given by
\begin{equation*}
D=\left[ \frac{3 \alpha -6}{3 \alpha -5}\right] \times \left[ \frac{W q f_M 5^{-\frac{\alpha}{2}}}{676 \Gamma \log 2}\right].
\end{equation*}
\label{theorem1}}
\end{Theorem}

When $d < \h$, bottlenecks form around the destinations, limiting the capacity of the network. Intuitively speaking, in this case there are so few destinations, that the convergence of traffic streams to each of them is so intense that the areas around them must carry many more traffic streams than other areas in the network. Therefore, each of these traffic streams must have a very small data rate, and this drives the whole capacity down. 

If, however, $\h < d < 1$, no bottlenecks are formed around the destinations, and the capacity can increase as fast as $n^\h$, as in the uniform traffic pattern case of \cite{gupta1}. Intuitively, the number of destinations is large enough so that, despite the asymmetry that still exists, the network can find a routing scheme that avoids congesting the areas around the destinations, and spreads the traffic evenly through the whole network. As the proof of the theorem will show, in order to achieve an aggregate throughput of $n^\h$, an average location in the network is required to support, on the average, $n^\h$ traffic streams. When $\h < d < 1$, the number $n^{1-d}$ of streams converging to a destination, which the location around the destination must support, is much less than that average load of $n^\h$ streams. Therefore, the extra workload of locations close to destinations is insignificant with respect to the average workload.

Although we do not formally prove the upper bound $n^\h$ on the capacity for the case $d > \h$, it is intuitively clear from the work in \cite{gupta1} that it holds, and so the lower bound is always tight up to a poly-logarithmic factor of the form $k_1 (\log n)^{k_2}$.

An important practical implication of Theorem~\ref{theorem1} is that networks can handle well \emph{some} asymmetry in the traffic pattern, but designers  should avoid any \emph{extreme} asymmetry. In particular, the number of destinations $m(n)$ should be at least on the order of $n^\h$, where $n$ is the number of sources. For applications in which $m(n)$ is a design parameter and it is useful to minimize it (because, for example, destinations are more expensive) the network has a `sweet-spot': $m(n)$ should be around $n^\h$. Using more destinations will not improve the performance significantly, but using fewer will severely reduce it. 

Before moving to the results for the other types of traffic, one clarification is needed regarding the selection of the number of destinations as $m(n)=n^d$. The result we provided holds for any $d \in A=(0,\frac{1}{2})\cup(\frac{1}{2},1)$. Therefore, the result allows us to scan a wide range of variations of $m$ with respect to $n$. Although we could have adopted a more general condition, such as $m \leq n$, we do not do so, because the additional derivations needed for addressing this more general case would be lengthy, without the value of the results increasing accordingly. In other words, our model is specific enough to keep the derivations in a manageable level, but general enough to provide intuition for all cases of interest. This discussion applies also to the other types of networks we study, for which similar assumptions are made.

\subsection{Multicast Networks}
 
\textbf{Multicast networks} consist of $n$ \textbf{wireless nodes} $X_1$, $X_2$, $\ldots$, $X_n$, placed randomly, and in particular uniformly and independently, in the area $\{(x,y): |x|,|y| \leq \h\}$. Each node creates traffic with a common rate $\lambda(n)$ that is intended for $m(n)=n^d$ other nodes, that are chosen randomly, uniformly and independently, among the rest. We call $d \in (0,1)$ the \textbf{multicast exponent}. Examples of networks with a multicast traffic pattern are wireless networks used in military or search-and-rescue operations where each user might want to communicate with an arbitrary subset of the other users. 

We define the \textbf{capacity $C(n)$ of the network} as the supremum of all rates $\lambda(n)$ that are uniformly achievable by all sources in the network, multiplied by their number $n$ \emph{and} the number of destinations $m(n)=n^d$. Note that the capacity is again a random variable.

\begin{Theorem} \emph{In multicast networks the capacity is bounded w. h. p. as follows: 
\begin{equation}
C(n) \geq \left[ \frac{3 \alpha -6}{3 \alpha -5}\right] \left[ \frac{W q f_M 5^{-\frac{\alpha}{2}}}{22000 \Gamma \log 2}\right] \frac{n^\frac{d+1}{2}}{(\log n)^\frac{3}{2}}.\label{eq3}
\end{equation}}
\label{theorem2}
\end{Theorem}

The improvement on the capacity over the uniform case is due to the possibility for the routing of each packet along a tree that passes through all destinations, as opposed to sending the same packet individually to each destination, in an uncoordinated manner. Although we formally present only a lower bound, we will use intuitive arguments to show that the routing tree employed by the constructive lower bound is of the same order of length as the minimum length multicast tree that the source can employ. For this reason, the lower bound is tight up to a poly-logarithmic factor.

An interesting side result is that the tight lower bound can be achieved without employing multicasting on the media access layer. The intuitive justification of this rather unexpected result is that any efficient multicast trees will have such a small number of bifurcations, so that employing multicasting in the media access layer cannot change the order of the capacity. Another interesting side result is that the tight lower bound can be achieved without the source discovering the location of the destinations, or the destinations discovering the location of the source. The only requirement is that each destination be discovered by a node carrying its packets that is on a distance at most $n^{-\frac{d}{2}}$ away from that destination.

\subsection{Cluster Networks}

\textbf{Cluster networks} consist of $n$ \textbf{client nodes} $X_1$, $X_2$, $\ldots$, $X_n$, and $m(n)=n^d$ \textbf{cluster heads} $Y_1$, $Y_2$, $\ldots$, $Y_m$, placed randomly, uniformly and independently, in the area $\{(x,y): |x|,|y| \leq \h\}$. We call $d \in (0,1)$ the \textbf{cluster head exponent}. Each client wants to establish a bidirectional communication (with rate $\lambda(n)$ in each direction) with  \emph{any} of the cluster heads. This model approximates well the traffic patterns that exist in wireless networks that operate using hierarchical clustering protocols, as for example Bluetooth~\cite{salonidis1}. Another application are sensor networks that consist of sensors and fusion centers.

We define the \textbf{capacity $C(n)$ of the network} as the supremum of all rates $\lambda(n)$ that are uniformly achievable by all data streams in the network, multiplied by their number $2n$. As in the previous cases, the capacity is a random variable. 

\begin{Theorem} \emph{In cluster networks the capacity is bounded w. h. p. as follows: 
\begin{eqnarray}
C(n) &\leq& \left[ \frac{4 \alpha W}{\log 2} \right] n^d \log n \label{eq4a}, \\
C(n) &\geq& \left[ \frac{W q f_M 5^{-\frac{\alpha}{2}}} {676 \Gamma \log 2} \right] \left[ \frac{3\alpha -6}{3 \alpha-5}\right] \frac{n^d}{(\log n)^2}. \label{eq4b}
\end{eqnarray}}
\label{theorem3}
\end{Theorem}

The theorem shows that, ignoring poly-logarithmic factors, the capacity increases with $n$ roughly as $n^d$. The upper bound (\ref{eq4a}) comes from the need of the network to share the area around the cluster heads. Therefore, the larger $d$ is, the faster the capacity increases with $n$. 

In the context of networks that use clustering, the theorem suggests that, to maximize capacity, the size of clusters must be bounded, and so their number should increase linearly with $n$. If network designers are not willing to accept such a large number of clusters, they should be ready to sacrifice part of the capacity. The exact tradeoff is very simple, and is captured by Theorem \ref{theorem3}. In the context of networks where the cluster heads are gateways to the outside world, the theorem suggests that there is no limit to how many gateways are needed: the greater the investment of the network provider (i.e., the larger $d$ is), the larger the capacity is going to be. Again, the tradeoff is very simple and is captured by Theorem~\ref{theorem3}.

Finally, as the proof will show, the lower bound on the capacity can be achieved even if clients do not transmit to each other, and even in the presence of fading (but in this last case, provided the client nodes are not restricted to communicate with the nearest cluster head). In other words, advanced routing protocols cannot change the capacity by more than a poly-logarithmic factor, and designers should focus instead on efficient polling algorithms that are aware of the channel state, and the efficient handling of bottlenecks around the cluster heads.

\subsection{Hybrid Networks}

\textbf{Hybrid networks} consist of $n$ \textbf{wireless nodes} $X_1$, $X_2$, $\ldots$, $X_n$, and $m(n)=n^d$ \textbf{access points} $Y_1$, $Y_2$, $\ldots$, $Y_m$, placed randomly, uniformly and independently, in the two-dimensional area $\{(x,y): |x|,|y| \leq \h\}$. We call $d \in (0,1)$ the \textbf{access point exponent}. We assume that the access points are connected with each other through a data link of infinite capacity that does not consume any of the available bandwidth $W$. There are $n$ traffic streams and each wireless node is the source of a single stream, and the destination of a single stream. A node cannot be the source and destination of the \emph{same} stream. Apart from this restriction, all other combinations of sources and destinations are equally probable. The access points do not have any communication needs of their own, but are there to support the wireless nodes. 

This network shares important common characteristics with both pure wireless multihop networks and also pure cellular networks: On the one hand, it partly consists of a large number of wireless nodes that communicate over a wireless channel and can route each other's traffic, as in wireless multihop networks. On the other hand, the wireless nodes are supported by access points that form an independent network with infinite capacity and do not have any traffic needs of their own; their role is similar to that of base stations in cellular networks.  The asymptotic capacity of such networks was first studied in~\cite{kozat1, towsley1}, and is of great practical interest, as future generation cellular systems will be using this hybrid topology. 

We define the \textbf{capacity $C(n)$ of the network} as the supremum of all rates $\lambda(n)$ that are uniformly achievable by all data streams in the network, multiplied by their number $n$. As in the previous cases, the capacity is a random variable.

\begin{Theorem}
\emph{In hybrid networks the capacity is bounded w. h. p. as follows:
\begin{equation}
C(n) \geq \h\left[ \frac{W q f_M 5^{-\frac{\alpha}{2}}} {676 \Gamma \log 2} \right] \left[ \frac{3\alpha -6}{3 \alpha-5}\right] \frac{n^d}{(\log n)^2},
\label{eq5}
\end{equation}
\begin{equation}
C(n) \geq \left[ \frac{3 \alpha -6}{3 \alpha -5}\right] \left[ \frac{W q f_m 5^{-\frac{\alpha}{2}}}{8600 \Gamma \log 2}\right]\frac{n^\frac{1}{2}}{(\log n)^\frac{3}{2}}.
\label{eq6}
\end{equation}}
\label{theorem4}
\end{Theorem}

Although we do not formally prove upper bounds, we provide an intuitive justification that (\ref{eq5}) is tight when $d>\h$, and (\ref{eq6}) is tight when $d<\h$.

The theorem suggests that more than $n^\h$ access points are needed for the infinite-capacity infrastructure to have any effect on the performance of the network. As the proof will reveal, no access point can expect to receive packets with a bit rate larger than $\log n$. Therefore, when $d<\h$, there are so few access point, so that even if they were receiving packets with that maximum possible rate, they would not be able to compete with the wireless network formed by the nodes, which can achieve an aggregate throughput on the order of $n^\h$.

If, however, $\h < d < 1$, there is a simple time division scheme, that does not depend on multihop wireless transmission, so that each wireless node can communicate with one of its neighboring wireless access nodes with rate $\frac{n^{d-1}}{(\log n)^2}$, which is much larger than $n^\h$. Therefore, the wireless nodes should not depend on each other for routing their traffic, but rather should make heavy use of the infrastructure. 

Note that there is a surprising phase transition: depending on how many access points there are, they should either be totally ignored, or used extensively. It is intuitively clear that the best strategy would be to use the full resources of both existing networks, however there will be no gain by doing this, \emph{in terms of the exponent} with which the aggregate throughput increases.

We note that a similar result was first reported in~\cite{kozat1, towsley1}. Our setup, however, is different in a number of critical ways: Firstly, we require that all wireless nodes are guaranteed the same throughput. Secondly, the locations on the access points are random, and finally we assume a more realistic channel model, that includes a general fading model. Our result is also straightforward to derive, because its proof is based on parts of the proofs of the other theorems presented in this work.

\section{Useful Lemmas}
\label{section:lemmas}

The first lemma is closely related to the well-known Coupon Collector's Problem~\cite{feller1}, however, to the best of our understanding, it has not appeared elsewhere in this form. 

\begin{Lemma}\emph{ Let $n$ balls be placed in $l$ urns, uniformly and independently of each other. Let $b_j$, $j=1,\ldots,l$ be the number of balls that end up in the $j$-th urn. Then for any $\epsilon>0$ there is a $\delta(\epsilon)>0$ such that $P[\forall j~(1-\epsilon) \frac{n}{l}\leq b_j \leq (1+\epsilon)\frac{n}{l} ] \geq 1-2l\exp[-\delta(\epsilon) \frac{n}{l}]$.}
\label{lemma1}
\end{Lemma}

\begin{Proof}{}
We make use of Chernoff's bounds~\cite{motwani}: Let $X$ be a binomial random variable, with parameters $k$ (the number of experiments) and $p$ (the probability of success of each experiment). For any $\epsilon >0$, 
\begin{eqnarray}
P[ X < (1-\epsilon) k p ] < \exp[-k p\frac{\epsilon^2}{2}], \label{eq7} \\
P[X>(1+\epsilon) k p] < \frac{\exp[\epsilon k p]}{(1+\epsilon)^{(1+\epsilon)k p}}
 \triangleq \exp[-k p f(\epsilon)],
\label{eq8}
\end{eqnarray}
where $f(\epsilon) \triangleq (1+\epsilon) \log(1+\epsilon) -\epsilon$. By calculating the derivative of $f(\epsilon)$ with respect to $\epsilon$, we have that $f(\epsilon)>0$ for $\epsilon>0$. 

Since each ball is placed in an urn independently of the others, $b_j$ follows the binomial distribution, with number of experiments equal to $n$ and probability of success equal to $\frac{1}{l}$. (Note, however, that the $b_j$ are not independent.) Applying Chernoff's bounds, we have:
\begin{eqnarray}
P[b_j<(1-\epsilon) \frac{n}{l}] &<& \exp[-\frac{\epsilon^2}{2}\frac{n}{l}], \label{eq9a} \\
P[b_j>(1+\epsilon)  \frac{n}{l}] &<& \exp[-f(\epsilon)\frac{n}{l}]. \label{eq9b}
\end{eqnarray}

We note the basic inequality $P[\cup_{j=1}^k E_j] \leq \sum_{j=1}^k P[E_j]$, typically referred to as the \emph{union bound}. Then:
\begin{eqnarray*}
 \lefteqn{P[\forall j~(1-\epsilon) \frac{n}{l}  \leq b_j \leq (1+\epsilon)\frac{n}{l} ]} \\
&=& 1- P[\forall j~(1-\epsilon) \frac{n}{l}  \leq b_j \leq (1+\epsilon)\frac{n}{l} ]^c \\
&\geq&  1 - \sum_{j=1}^l \left\{ P [b_j < (1-\epsilon) \frac{n}{l}] + P [b_j > (1+\epsilon)\frac{n}{l}] \right\} \\
 &\geq&  1- l \left\{ \exp[-\frac{\epsilon^2}{2} \frac{n}{l}]+  \exp[-f(\epsilon)\frac{n}{l}] \right\} \\
 &\geq&  1-2l\exp[-\delta(\epsilon)\frac{n}{l}],
\end{eqnarray*}
where $\delta(\epsilon) \triangleq \min\{\frac{\epsilon^2}{2},f(\epsilon)\}>0$. The first inequality comes from the union bound, and the second inequality from (\ref{eq9a}) and (\ref{eq9b}).
\end{Proof} 

In subsequent sections, we will have to bound the effects of interfering transmissions in the reception of signals. As the fading distribution has an exponentially thin tail, the following lemma applies:  

\begin{Lemma}
\emph{Let $n$ nodes communicating over a wireless channel that satisfies the assumptions set out in Section~\ref{section:model}. With high probability, the maximum value of fading coefficients between all pairs of nodes is bounded as follows:
\begin{equation*}
\max_{1 \leq i < j \leq n} \{f_{ij} \} \leq \frac{3}{q} \log n.
\end{equation*}
\label{lemma2}}
\end{Lemma}

\begin{Proof}{}
Let the events $F_{ij}(n) \triangleq \{ f_{ij}> \frac{3}{q} \log n\}$. Then:
\begin{multline*}
P\left[\max_{1 \leq i < j \leq n} \{f_{ij} \} \leq \frac{3}{q} \log n \right]= 1-P[\cup_{1 \leq i < j \leq n}F_{ij}(n)] \\
\geq  1 - \sum_{1 \leq i < j \leq n} P[F_{ij}(n)]
 \geq  1-\frac{n(n-1)}{2} n^{-3} 
\rightarrow  1.
\end{multline*}
The first inequality comes from the union bound. The second comes from symmetry and applying (\ref{eq1}), and holds only for sufficiently high $n$. 
\end{Proof}

Finally, observe that if a sequence of events $A_n$ occurs w. h. p., and a second sequence of events $B_n$ occurs w. h. p. conditioned on the sequence $A_n$, then $B_n$ will also occur w. h. p. without the conditioning: 
\begin{Lemma}
\emph{Let $\underset{n \rightarrow \infty}{\lim}P[A_n]= 1$ and $\underset{n \rightarrow \infty}{\lim} P[B_n|A_n] =1$. Then  $\underset{n \rightarrow \infty}{\lim}P[B_n] = 1$.}
\label{lemma3}
\end{Lemma}
The proof follows immediately by noting that $P[B_n]= P[B_n|A_n]P[A_n] +P[B_n|A_n^c]P[A_n^c]$. In practical terms, if we need to prove that a sequence of events occurs w. h. p., we are free to condition the discussion on any sequence of events that occurs also w. h. p. It is also clear that we can iteratively condition on more than one sequence of events. We will use this lemma repeatedly, in many cases implicitly.

\section{Asymmetric Networks}
\label{section:asymmetric}

We first develop a constructive proof for the lower bound (\ref{eq2b}) of Theorem~\ref{theorem1} in the spirit of \cite{gupta1}: we develop a communications scheme whose aggregate throughput equals the lower bound w. h. p., and as the capacity is the supremum of the aggregate throughputs of \emph{all} schemes, it will necessarily exceed this lower bound.

\subsection{Cell Lattice}
\label{subsection:lattice}

\begin{figure}
\begin{center}
\includegraphics[scale=0.40]{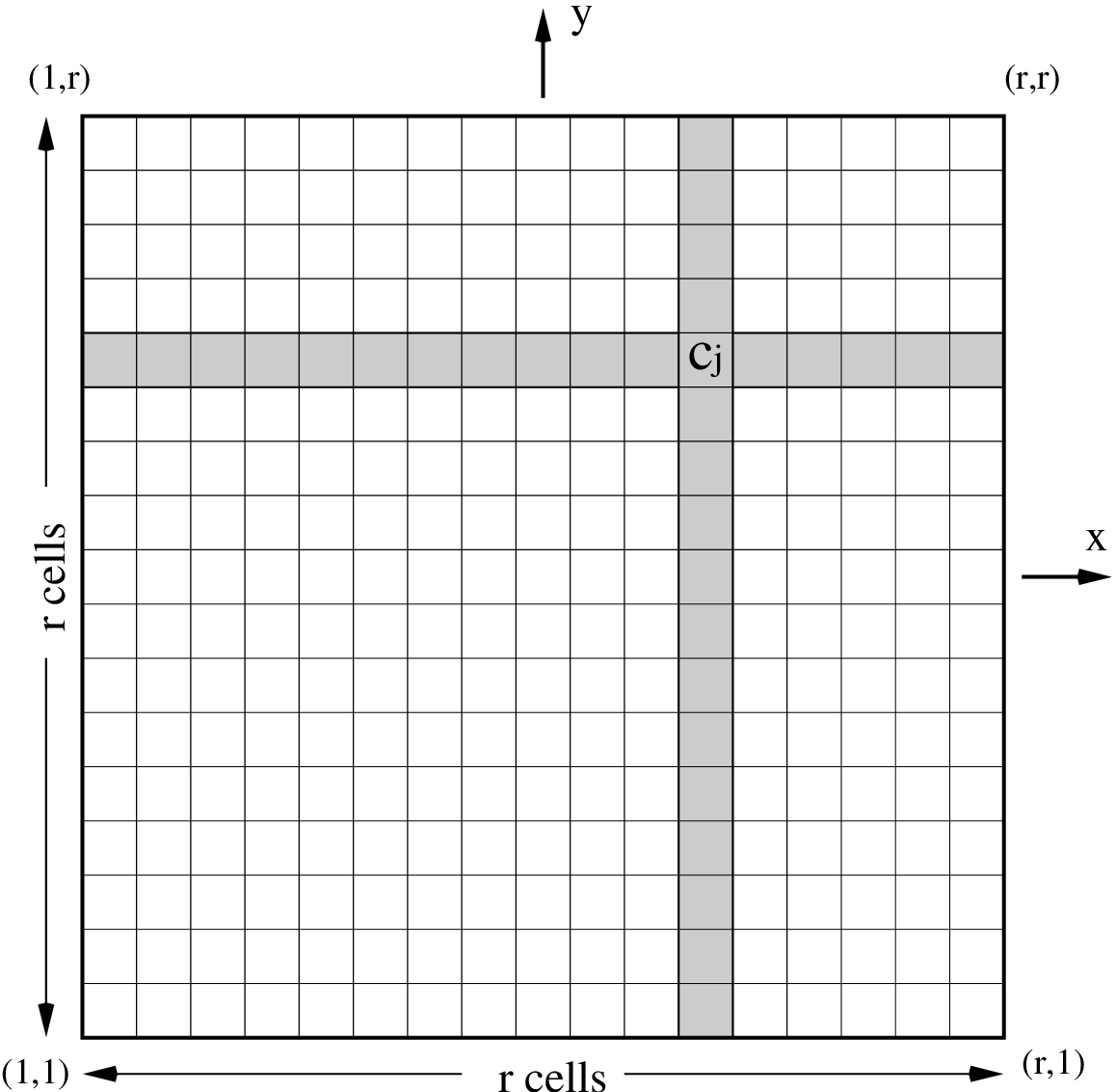}
\end{center}
\caption{Partition of the square region $\{(x,y): |x|,|y| \leq \h\}$ into a regular lattice of $r^2$ cells. We define $s_j$ as the number of source nodes in cell $c_j$, $M_j$ as the number of \emph{source} nodes lying in cells who share the same x-coordinate with $c_j$ (the shaded cell column) and $N_j$ as the number of \emph{destination} nodes lying in cells who share the same y-coordinate with $c_j$ (the shaded cell row).}
\label{figure:cells}
\end{figure}

As shown in Fig.~\ref{figure:cells}, we divide the square region $\{(x,y): |x|,|y| \leq \h\}$ in a regular lattice of $g(n)=\frac{n}{18\log n} \triangleq r^2$ cells $c_1,~c_2,~\ldots,c_{g(n)}$. Each cell can be identified by its coordinates $(v_1,v_2)$ in the lattice, where $ 1 \leq v_1,~v_2 \leq r $; the cell on the lower left corner has coordinates $(1,1)$. We call two cells \textbf{neighbors} if they share a common boundary edge, so that each cell has at most four neighbors.

Let $s_j$ be the number of source nodes in cell $c_j$. Thinking of cells like urns and source nodes like balls, we see that Lemma~\ref{lemma1} applies. Setting $\epsilon=\frac{1}{2}$, $l=g(n)$, $b_j=s_j$, $\delta(\epsilon)= \min\{\frac{\epsilon^2}{2}, (1+\epsilon)\log(1+\epsilon)-\epsilon \}>\frac{1}{10}$, it follows that $P\left[\forall j~9\log n \leq s_j \leq 27 \log n \right] \geq 1-\frac{2n^{-\frac{8}{10}}}{18\log n}$, which goes to $1$ as $n \rightarrow \infty$. Therefore, w. h. p.,
\begin{equation}
\forall j,~~9\log n \leq s_j \leq 27 \log n.
\label{eq10}
\end{equation}

Note that we could have selected a different value for $\epsilon$ within $(0,1)$; the critical requirement is to show that w. h. p. $s_j$ equals $\log n$, up to at most a constant factor. Next, let $F_{ij}(n)$ be the event that the source node $X_i$ cannot find a source node in one of its neighboring cells $c_j$, such that their mutual fading coefficient is greater or equal to $f_M$. By the independence of the fading coefficients, and using (\ref{eq10}), it follows that $P[F_{ij}(n)] \leq (\frac{1}{2})^{9 \log n}$. By using the union bound, and noting that there are $n$ source nodes, each with at most $4$ neighboring cells, it follows that $P[\cup_{i,j} F_{ij}] \leq 4n(\frac{1}{2})^{9 \log n} \rightarrow 0$. Therefore, w. h. p. each source node will be able to find another source node in each of the neighboring cells, such that their mutual fading coefficient is equal to or greater than $f_M$. 

Finally, let $G_{ij}(n)$ be the event that a destination node $Y_i$ and a source node $X_j$ lying in the same cell will not be able to find a relaying source node $X_k$, also on that cell, such that the mutual fading coefficients $f_{Y_i X_k} \geq f_M$ and $f_{X_k X_j} \geq f_M$. By the independence of the fading coefficients, the probability that a particular source node cannot be used is at most $\frac{3}{4}$, and the probability that there is no source node that can be used is at most $(\frac{3}{4})^{9\log n}$. Applying the union bound, it follows that the probability $P[\cup_{i,j} G_{ij}(n)] \leq n^d (27 \log n) (\frac{3}{4})^{9\log n} \rightarrow 0$. Therefore, w. h. p. any destination node will be able to communicate with any source node in its cell, by using another source node in that cell as a relay, and in both hops the fading coefficient will be greater or equal to the median $f_M$. 

Let us summarize the results until now: We have divided our area into $\frac{n}{18\log n}$ cells and we have shown that the following properties hold w. h. p.: \textbf{(i)} The numbers of source nodes in all cells are bounded by (\ref{eq10}). \textbf{(ii)} Each source node can find a source node in any of its neighboring cells so that their mutual fading coefficient is greater than or equal to the median $f_M$. \textbf{(iii)} Each source node can communicate with any of the destination nodes in its cell through a relaying source node in that cell, so that the fading coefficients of both hops are greater than or equal to the median $f_M$. From now on, we condition the discussion on the assumption that these three results hold. By Lemma~\ref{lemma3}, if a property such as a capacity bound holds w. h. p. conditioned on these results, it will also hold w. h. p. without the conditioning.

\subsection{Routing Protocol}
\label{subsection:routing_protocol}

\begin{figure}
\begin{center}
\includegraphics[scale=0.50]{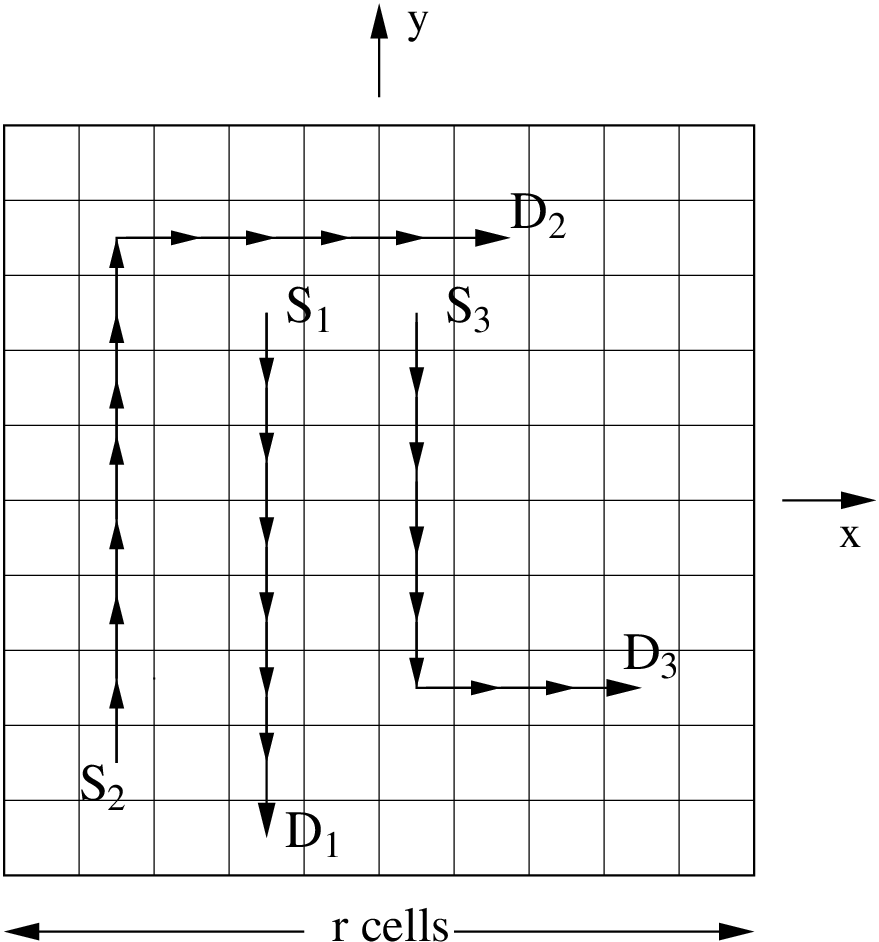}
\end{center}
\caption{Examples of routes used in asymmetric networks.}
\label{figure:routes}
\end{figure}

As shown in Fig.~\ref{figure:routes}, packets are routed according to the following rules:
\begin{list}{}{\setlength{\leftmargin}{0pt}\setlength{\itemsep}{0pt}}
\item[]\textbf{(i)} If a source node $X_j$ has data packets (possibly not created at $X_j$) that must be delivered to a destination node $Y_i$ lying in the same cell, and $f_{X_j Y_i} < f_M$, $X_j$ will transmit the data packets to another source node $X_k$ lying in the same cell, for which $f_{X_j X_k} \geq f_M$ and $f_{X_k Y_i} \geq f_M$. Node $X_k$ will then transmit the packet to the destination node $Y_i$.  By the discussion of Section~\ref{subsection:lattice}, we can assume that such a node exists. 
\item[]\textbf{(ii)} If the destination node $Y_j$ of a source node $X_i$ lies in a different cell from $X_i$, the packets of $X_i$ are routed through intermediate cells. In particular, only communication between source nodes who lie in neighboring cells and whose mutual fading coefficient is at least equal to the median is allowed. In addition, the packets are first transmitted along cells whose x-coordinate is the same as the x-coordinate of the source, until they arrive at a cell whose y-coordinate is the same as the y-coordinate of the destination. Then, the packets are transmitted along cells whose y-coordinate is the same as the y-coordinate of the destination, until they arrive at a source node lying in the same cell with the destination. By the discussion of Section~\ref{subsection:lattice}, we can assume that such relays always exist. Once the packets arrive at the cell of the destination, they are delivered to the destination as specified by rule \textbf{(i)}.
\end{list}

To evaluate the performance of this scheme, we must calculate the load that the routing protocol creates for each cell. To this end, let us define $M_j$ as the number of source nodes that lie in cells whose $x$-coordinate is the same as the $x$-coordinate of cell $c_j$, and $N_j$ as the number of destination nodes that lie in cells whose $y$-coordinate is the same as the $y$-coordinate of cell $c_j$. We develop bounds on the values of $M_j$ and $N_j$ that we will use to bound the traffic that each cell must support. 

To bound the value of $M_j$, we note that there are $\sqrt{\frac{n}{18 \log n}}$ cells with the same $x$-coordinate with cell $c_j$, each with at most $27\log n$ source nodes. Therefore:
\begin{equation}
\forall j,~M_j \leq \sqrt{\frac{n}{18 \log n}}(27 \log n) =\frac{9}{\sqrt{2}} \sqrt{n \log n}.
\label{eq11}
\end{equation}

Next, we bound $N_j$ for the case $d>\frac{1}{2}$. Applying Lemma~\ref{lemma1} with $\epsilon=\frac{1}{2}$ where the balls are the destination nodes and the urns are the rows:
\begin{multline*}
P[\forall j~ \frac{3}{\sqrt{2}}n^{d-\h}\sqrt{\log n} \leq N_j \leq \frac{9}{\sqrt{2}}n^{d-\h}\sqrt{\log n} ] \\
\geq 1-2 \sqrt{\frac{n}{18 \log n}} \exp[-\delta(\frac{1}{2}) 3\sqrt{2} n^{d-\h}\sqrt{\log n}] \rightarrow 1.
\end{multline*}
By Lemma~\ref{lemma3}, we are allowed to assume that:
\begin{equation}
d>\h \Rightarrow \forall j,~N_j \leq \frac{9}{\sqrt{2}}n^{d-\h}\sqrt{\log n}.
\label{eq12}
\end{equation}

Finally, we uniformly bound the $N_j$ for the case $d<\h$. For this we use (\ref{eq8}), noting that $N_j$ follows the binomial distribution with $p=(\frac{n}{18\log n})^{-\h}$ and $k=n^d$. Setting $\epsilon$ to satisfy $(1+\epsilon)k p=x$, where $x$ will be specified later, we have that: 
\begin{multline*}
 P[N_j>x]
< \exp[x-k p](\frac{k p}{x})^x \\
< \frac{\exp[x]}{x^x} \left( 3\sqrt{2} n^{d-\h} \sqrt{\log n} \right)^x.
\end{multline*}
Applying the union bound, we have that $P[\exists j:~N_j>x] \leq \frac{\exp[x]}{x^x} \left( 3\sqrt{2} n^{d-\h} \sqrt{\log n} \right)^x \sqrt{\frac{n}{18\log n}}$, which goes to $0$ if we choose $x>\frac{1}{1-2d}$, for example $x=\frac{2}{1-2d}$. Applying Lemma \ref{lemma3}, we can assume that:
\begin{equation}
d<\h \Rightarrow \forall j,~N_j\leq \frac{2}{1-2d}.
\label{eq13}
\end{equation}

\begin{Lemma} \emph{Let $r_j$ be the number of routes arriving, and possibly terminating, at cell $c_j$. Then w. h. p.:
\begin{equation*}
\forall~j,~r_j \leq r_{\mathrm{max}}(n) \triangleq \begin{cases} \frac{27}{\sqrt{2}} (n \log n)^\h & \mbox{if }~ \h < d < 1, \\
\frac{5}{1-2d}n^{1-d} & \mbox{if }~ 0 < d < \h.
\end{cases}
\end{equation*}}
\label{lemma7}
\end{Lemma}
\begin{Proof}{}
Let $r_{j1}$ be the number of routes that cross $c_j$ while on their vertical leg (see Fig.~\ref{figure:routes}). The sources of those routes share a common $x$-coordinate with $c_j$. Also, let $r_{j2}$ be the number of routes that cross $c_j$ while on their horizontal leg. The destination nodes of these routes share a common $y$-coordinate with $c_j$. Each route crossing $c_j$ will belong to one or both of the two types of routes, so necessarily $r_j \leq r_{j1}+r_{j2}$. Therefore, it suffices to bound both $r_{j1}$ and $r_{j2}$ uniformly for all cells $c_j$.

As each source node is the source of a single stream, $r_{j1} \leq M_j$. To bound $r_{j2}$, we note that, by a straightforward application of Lemma~\ref{lemma1}, at most $2n^{1-d}$ routes can be terminating at each destination, w. h. p. Therefore $r_{j2} \leq 2n^{1-d} N_j$ w. h. p. Combining these inequalities we have that $r_j \leq M_j + 2 n^{1-d}N_j$ w. h. p., for all cells $c_j$.  The result follows by using (\ref{eq11}), (\ref{eq12}), and (\ref{eq13}), also noting that when $d<\h$, $\frac{\sqrt{n \log n}}{n^{1-d}} \rightarrow 0$.
\end{Proof}

Since there are $n$ routes, each requiring a number of hops on the order of $(\frac{n}{\log n})^\frac{1}{2}$, and the total number of hops must be shared by $\frac{n}{18\log n}$ cells, on the average each cell will be required to relay a number of routes on the order of $(n \log n)^\frac{1}{2}$. Therefore, Lemma~\ref{lemma7} implies that when $d > \h$, no cell will have to carry much more that its `fair share' of the traffic. If, however, $d < \h$, then there are so few destinations, that a few `unlucky' cells (those on the same column with a destination) will be required to serve around $n^{1-d}$ routes, which is much more than their `fair share' of traffic. In those cells, bottlenecks will form.

\subsection{Time Division}
\label{subsection:time_division}

Until now, we have specified a routing protocol, based on cells, provided guarantees that communication will be between nodes that are not in deep fades, and proved bounds on the amount of traffic that each cell will need to support. However, we have not specified a medium access protocol. Such a medium access protocol is needed so that each cell knows when to transmit, and also there are guarantees about the minimum amount of traffic that each cell can support. In this section, we develop such a medium access protocol, based on time division.

We divide the $g(n)=r^2$ cells into nine regular sub-lattices, such that any two cells belonging in the same sub-lattice are separated by at least two cells belonging to different sub-lattices. This property will be used to bound the amount of interference experienced by receivers. In Fig.~\ref{figure:interference} we have shaded the cells belonging to one of the $9$ sub-lattices.

We divide time into frames, and each frame into nine slots, each slot
corresponding to a sub-lattice. At any time during that slot, only one node from each cell of the corresponding sub-lattice is allowed to \emph{receive} (but many nodes in that cell may receive consecutively in the same slot). Because of the way we constructed the routing protocol, the transmitter of that transmission will have to lie in the same cell, or in one of the four neighboring cells. All transmissions are with the maximum power $P_0$.

\begin{Lemma} \emph{The SINR $\gamma_j$ at any source or destination node $Z_j$ that is receiving is bounded w. h. p. by
\begin{equation}
\gamma_j > \gamma_{\mathrm{min}}(n) \triangleq  5^{-\frac{\alpha}{2}} \left[\frac{ 3\alpha-6}{3\alpha-5}\right] \left[\frac{q f_M}{25}\right] \frac{1}{\log n}.
\label{eq14}
\end{equation}}
\label{lemma8}
\end{Lemma}

\begin{Proof}{}
We first bound the interference $I_j$. For this, we first note that by Lemma~\ref{lemma2}, w. h. p. no fading coefficient is greater than $\frac{3}{q} \log n$. Next, let $x_0=\frac{1}{r}$ be the length of the sides of the cells, and let $c_k$ be the cell in which the receiving node lies. Working as in~\cite{bansal1}, we note that the rest of the cells in the same sub-lattice are located along the perimeters of concentric squares, whose center is cell $c_k$. Irrespective of the coordinates of $c_k$, all the cells of its sub-lattice are located along the perimeters of at most $\f{\frac{r-1}{3}}$ squares. There are at most $8i$ interferers corresponding to the $i$-th square, whose distances from the receiver will be at least $x_0 (3i-2)$. Consequently, the interference at the receiver is upper bounded by
\begin{eqnarray}
I_j & \leq & \left[\frac{3}{q} \log n \right] \sum_{i=1}^{\f{\frac{r-1}{3}}} \frac{8iKP_0}{[x_0(3i-2)]^\alpha}\nonumber \\
 &\leq&  \left[\frac{3}{q} \log n \right] \frac{8KP_0}{x_0^\alpha} ~[ 1 + \sum_{i=2}^{r} (3i-2)^{1 -\alpha} ]\nonumber \\
& < & \left[\frac{3}{q} \log n \right] \frac{8KP_0}{x_0^\alpha} ~[ 1+ \int_0^{r} (3x+1)^{1-\alpha}~d x] \nonumber \\
 &\leq& \left[\frac{3}{q} \log n \right] \frac{8KP_0}{x_0^\alpha} \left[\frac{3\alpha-5}{3\alpha-6}\right].  \label{eq15}
\end{eqnarray}

\begin{figure}
\begin{center}
\includegraphics[scale=0.40]{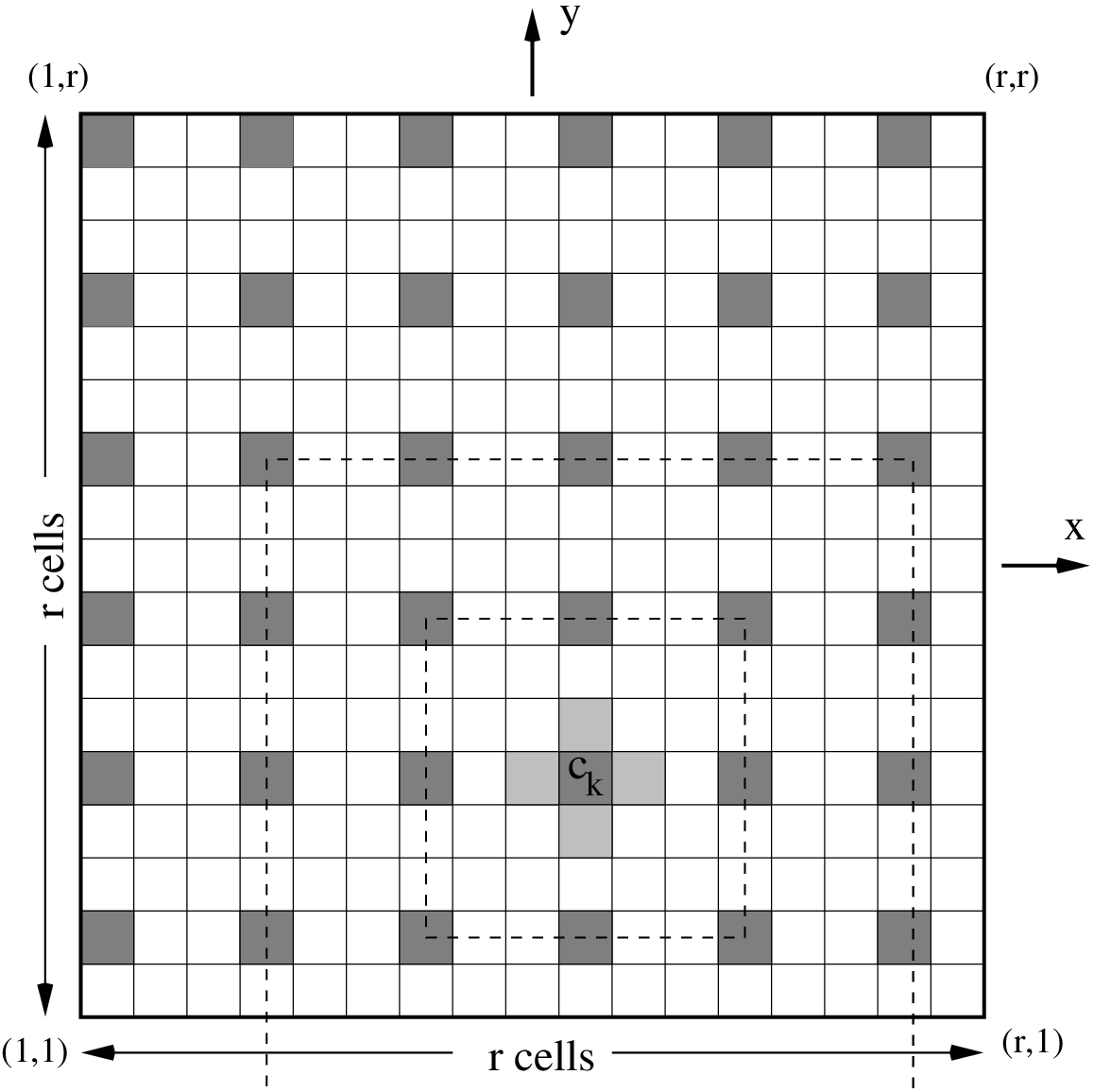}
\end{center}
\caption{One of the $9$ sub-lattices of cells appears shaded. Only nodes in that sub-lattice are allowed to \emph{receive} in the corresponding slot, and only from nodes in the same or neighboring cells. The neighbors of cell $c_k$ are lightly shaded. The cells belonging to the same sub-lattice as cell $c_k$ may be placed in at most $\f{\frac{r-1}{3}}$ concentric squares of increasing size, centered at $c_k$. The first two such squares are denoted by dashed lines.}
\label{figure:interference}
\end{figure}

We also need a lower bound on the power of the useful signal. Clearly, since the maximum possible distance that the useful signal will need to travel, under the routing assumptions, is $\sqrt{5} x_0$, and the fading coefficient between the transmitter and the receiver is at least equal to $f_M$, w. h. p. we have that $S_j \geq K P_0 f_M (\sqrt{5}x_0)^{-\alpha}$. Combining this with (\ref{eq15}), and noting that the thermal noise remains bounded, and therefore becomes negligible as $n \rightarrow \infty$, we arrive at (\ref{eq14}).
\end{Proof}

We now assume that all transmitters transmit with rate $f_R(\gamma_{\mathrm{min}}(n))$. By Lemma~\ref{lemma8}, w. h. p. all transmissions will be successful.

\subsection{Lower Bound}
\label{subsection:lower_bound}

The nodes of each cell are allowed to receive during only $1$ out of $9$ slots, and with rate equal to $f_R(\gamma_{\mathrm{min}}(n))$. The number of routes that will be crossing each cell $c_j$ is upper bounded by $r_{\mathrm{max}}(n)$, determined by Lemma~\ref{lemma7}. Most of these routes will require one reception, however a few of these, in particular those whose destination lies in cell $c_j$, may require three receptions. Therefore, each route, and its associated source node, is guaranteed a rate of communication $\lambda(n)= f_R(\gamma_{\mathrm{min}}) \left[ 3 \times 9 \times r_{\mathrm{max}}(n)\right]^{-1}$. Multiplying by $n$, and substituting for $r_{\mathrm{max}}(n)$ and $\gamma_{\mathrm{min}}(n)$ from Lemmas~\ref{lemma7} and~\ref{lemma8} respectively, we see that our scheme achieves an aggregate throughput equal to the lower bound (\ref{eq2b}). Since the capacity is the supremum of the aggregate throughputs of \emph{all} possible schemes, it will necessarily be greater than the aggregate throughput of our scheme, and the result follows.

\subsection{Proof of Upper Bound}
\label{subsection:upper_bound}

Let $d_{\mathrm{min}}$ be the minimum of all distances between all $m n$ source-destination pairs, and let $H_{ij}(x)$ be the event $\{|X_i-Y_j|\leq x\}$. Then:
\begin{multline}
P[d_{\mathrm{min}} \leq x]  =  P[\cup_{i,j}H_{ij}(x)] 
\leq  \sum_{i=1}^n \sum_{j=1}^m  P[H_{ij}(x)]\\
\\ =  n m  P[H_{11}(x)]
 \leq  n m \pi x^2. 
\label{eq16}
\end{multline}
The first inequality comes from the union bound. The second equality comes from using symmetry. The last inequality comes from noting that the nodes are placed in a square with surface area equal to $1$, and that nodes $X_1$ and $Y_1$ will be within distance $x$ of each other if $Y_1$ is placed on the intersection of the square with a disk of radius $x$, centered at node $X_1$.

The capacity is less than the aggregate throughput $T(n)$ that would have been achieved if all destination nodes were receiving \emph{(i)} all the time, \emph{(ii)} from sources at the minimum distance $d_{\mathrm{min}}$ and with fading coefficient equal to the upper bound of Lemma~\ref{lemma2}, \emph{(iii)} using the whole bandwidth, and \emph{(iv)} without experiencing interference from competing transmissions. Therefore, we can bound $T(n)$ as follows:
\begin{multline*}
T(n) \leq  m W \log_2(1+\frac{1}{\Gamma} \frac{KP_0 d_{\mathrm{min}} ^{-\alpha} \frac{3}{q} \log n}{\eta}) \\
 \leq   m W \log_2(1+\frac{3KP_0}{\eta q\Gamma} n^{3\alpha}  \log n)
 \leq   \left[ \frac{4 \alpha W}{\log 2} \right] n^d \log n. 
\end{multline*}
The second inequality holds w. h. p., and comes by applying (\ref{eq16}) with $x=n^{-3}$. The last holds for sufficiently large values of $n$, and comes using simple properties of the logarithm function. Since $C(n) \leq T(n)$, the bound follows.

\section{Multicast Networks}
\label{section:multicast}

In order to better motivate the proof of the lower bound of Theorem~\ref{theorem2}, we first present a heuristic upper bound\footnote{We note that a  similar bound, based on a different heuristic argument, appeared in the independent work in \cite{jacquet2}.}: to minimize the number of transmissions needed for a packet to reach all its destinations, it is clear that the packet must be routed along a \textbf{multicast tree} which passes through all the destinations and has as small a length as possible. Let us find a lower bound on the length of \emph{any} tree that connects all destinations. For this, let us divide the whole region in $q(n)=\frac{n^d}{18\log n}$ regular cells, each with side length equal to $q(n)^{-\frac{1}{2}}$. Working as Section~\ref{section:asymmetric}, it follows that there will be a destination in each of them, so a tree connecting all of them will have a length on the order of $q(n) \times q(n)^{-\frac{1}{2}}=q(n)^\frac{1}{2}\simeq n^\frac{d}{2}$, ignoring logarithmic factors. Assuming transmissions across distances which are as small as possible, i.e., on the order of $n^{-\frac{1}{2}}$, it follows that each packet will need roughly $n^\frac{d+1}{2}$ transmissions to be delivered to all $n^d$ destinations. As the number of simultaneous transmissions (across distances on the order of $n^{-\h}$) over the whole network is on the order of $n$ (using, for example the time division scheme of the previous section), it follows that the maximum possible aggregate rate of packet deliveries at destinations is on the order of $n \times n^d \times [n^\frac{d+1}{2}]^{-1}=n^\frac{d+1}{2}$. Up to the exponent of $n$, this upper bound equals the lower bound we now derive. 

Moving to the constructive proof of the lower bound, ideally, we would like to construct a scheme that uses a multicast tree that is as short as possible, for example a Steiner tree on a properly defined graph. However, we also need a tree that is amenable to analysis. The tree we now specify represents a good compromise between these goals.

\begin{figure}
\begin{center}
\includegraphics[scale=0.55]{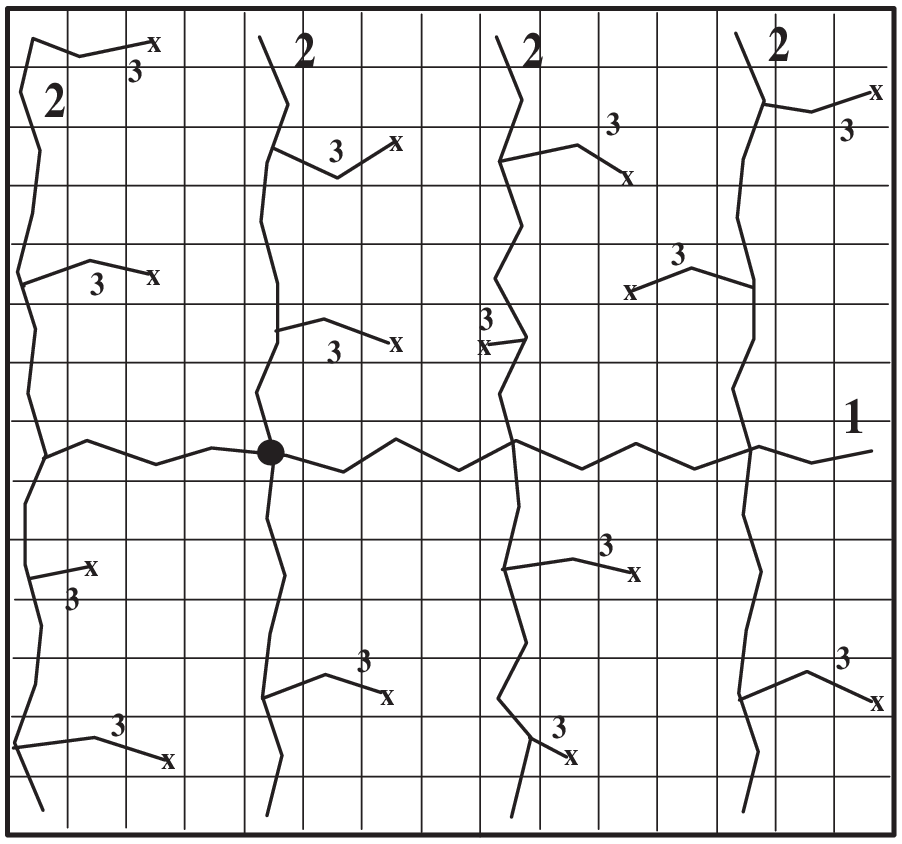}
\end{center}
\caption{An example of a multicast tree created according to the rules of Section~\ref{section:multicast}. The three legs of the tree are denoted by the numbers $1$, $2$, $3$. The source is denoted by a full circle and the destinations by crosses. Relaying nodes are not shown.}
\label{figure:simple_route}
\end{figure}

First, we divide the square region $\{(x,y): |x|,|y| \leq \h\}$ into $g(n)=\frac{n}{18 \log n}$ cells. The properties \textbf{(i)}-\textbf{(iii)} of Section \ref{subsection:lattice} continue to hold, where now property \textbf{(iii)} applies to the communication of two wireless nodes in the same cell. As shown in Fig.~\ref{figure:simple_route}, the tree we use consists of three legs:
\begin{list}{}{\setlength{\leftmargin}{0pt}\setlength{\itemsep}{0pt}}
\item{}\textbf{First leg:} The packet is propagated along a straight line to all cells which have the same $y$-coordinate as the cell of the source. 
\item{}\textbf{Second leg:} Starting from the cell of the source, every $h(n) \triangleq \frac{n^\frac{1-d}{2}}{3 \sqrt{ \log n}}$ cells along the first leg, the packet also propagates along the vertical direction. Therefore, there are $\frac{n^\frac{d}{2}}{\sqrt{2}}$ vertical legs per tree, separated by a distance of $n^{-\frac{d}{2}}$.
\item{} \textbf{Third leg:} Each destination receives the packet from the cell that received the packet in the second leg which is closest.
\end{list}
As with the routing protocol of Section~\ref{subsection:routing_protocol}, communication is between nodes those mutual fading coefficient is no smaller than the median $f_M$. Also, if a packet reaches a node in the cell of the destination other than the destination, it will reach the destination by two more hops, through a relay node, such that both mutual fading coefficients are at least equal to the median. 

The aim of the first two legs is to spread the packet uniformly through the whole region, and the number of vertical sections strikes the optimal balance between having a small number of total hops and a thick coverage. Ignoring poly-logarithmic factors and the fact that packets do not follow exactly straight lines, we note that the length of the tree is $1+n^\frac{d}{2}\times 1 + n^d \times n^{-\frac{d}{2}} \simeq n^\frac{d}{2}$. Therefore, this tree has the potential to achieve our heuristic upper bound, at least up to a poly-logarithmic factor.

Next, we develop an upper bound on the traffic supported by each cell. For this, let $r_j$ be the number of routes that cell $c_j$ must support, and let $r_{j1}$, $r_{j2}$, and $r_{j3}$ be the total number of routes passing through $c_j$ in their first, second, and third leg respectively. Clearly, $r_j \leq r_{j1}+r_{j2}+r_{j3}$. 

To bound $r_{j1}$, we apply Lemma~\ref{lemma1} as in the case of the bound (\ref{eq11}) and conclude that, for all $j$, $r_{j1} \leq \frac{9}{\sqrt{2}}\sqrt{n \log n}$. To bound $r_{j2}$, we note that each of the $n$ nodes will contribute to $r_{j2}$ with one route with probability $h(n)^{-1}$, and so by a simple application of the Chernoff and union bounds, w. h. p., for all $j$, $r_{j2} \leq \frac{3}{2}h(n)^{-1}n =\frac{9}{2} n^\frac{1+d}{2} \sqrt{\log n}$. To bound $r_{j3}$, we note that a cell $c_j$ will only have to serve \emph{some} of the third legs of routes with destinations that lie in either $c_j$, or in one of the $h(n)$ cells on its left, or in one of the $h(n)$ cells on its right. Therefore, $r_{j3}$ is at most equal to the number of destinations in $2h(n)+1$ cells. By Lemma \ref{lemma1}, w. h. p. there are at most $27 \log n$ nodes in each of these cells, for a total of at most $[2h(n)+1] 27 \log n$ nodes. The probability that one of these nodes is chosen when a node chooses his next destination is $[2h(n)+1] 27 (\log n)n^{-1}$. Applying the Chernoff bound (\ref{eq8}) with number of experiments $n^{1+d}$ and probability of success $[2h(n)+1] 27 (\log n)n^{-1}$, it follows that $r_{j3} \leq 27 n^{\frac{1+d}{2}} \sqrt{\log n}$, with probability going to $1$ exponentially fast. By a simple application of the union bound, it follows that w. h. p. the inequality will hold for all $j$. Combining the bounds for $r_{j1}$, $r_{j2}$, $r_{j3}$, it follows that w. h. p., and for all $j$, 
\begin{multline}
r_j \leq r_{j1}+r_{j2}+r_{j3} \leq  \frac{9}{\sqrt{2}}\sqrt{n \log n} + \frac{9}{2} n^\frac{1+d}{2} \sqrt{\log n} \\ +  27 n^{\frac{1+d}{2}} \sqrt{\log n} \leq r_{\mathrm{max}}(n) \triangleq 32n^{\frac{1+d}{2}}(\log n)^\frac{1}{2}.
\label{eq17}
\end{multline}

Next, we specify that the nodes use the time division schedule of Section~\ref{subsection:time_division}, under which each receiver is guaranteed, w. h. p., an SINR equal to the bound $\gamma_{\mathrm{min}}(n)$ given by (\ref{eq14}). Also, every transmitter transmits with rate $f_R(\gamma_{ \mathrm{min}}(n))$. The number of routes crossing each cell is at most $r_{\mathrm{max}}(n)$, given by (\ref{eq17}). Most of these involve just one hop, however those few whose destination lies in cell $c_j$ will require three transmissions. Therefore, each route is guaranteed a rate of communication 
$\lambda(n)= f_R(\gamma_{\mathrm{min}}(n)) \left[ 3 \times 9 \times r_{\mathrm{max}}(n)\right]^{-1}$. Multiplying with $n$, for the number of nodes, and $n^d$, for the number of destinations of each node, we arrive that the lower bound (\ref{eq3}).

\section{Cluster Networks}
\label{section:cluster}

Regarding the upper bound (\ref{eq4a}) of Theorem~\ref{theorem3}, we simply note that we can prove it by applying the technique used for proving the upper bound (\ref{eq2a}): we must simply consider upper bounds on the aggregate throughput received at the \emph{cluster heads}, as opposed to the \emph{destination nodes}. 

We next present a constructive proof of the lower bound (\ref{eq4b}). We divide the square region $\{(x,y): |x|,|y| \leq \h\}$ in a regular lattice of $q(n)=\frac{n^d}{18 \log n} \triangleq r^2$ cells, as shown in Fig.~\ref{figure:cells}. Let $s_j$ and $d_j$ be the numbers of client nodes and cluster heads respectively in cell $c_j$. By Lemma~\ref{lemma1}, it follows that w. h. p. 
\begin{eqnarray}
\forall j,~~~ 9n^{1-d}\log n \leq s_j \leq 27 n^{1-d} \log n, \label{eq20}\\
\forall j,~~~ 9\log n \leq d_j \leq 27\log n.  \label{eq21}
\end{eqnarray}

The probability that a client will not be able to find a cluster head in its own cell such that their mutual fading coefficient is greater than the median $f_M$ is, using the independence of different fading coefficients, at most $(\frac{1}{2})^{9\log n}$. Using the union bound, it follows that the probability that \emph{any} of the clients will not be able to find such a cluster head is smaller than $n(\frac{1}{2}) ^{9\log n}$, which converges to $0$ as $n \rightarrow \infty$. Therefore, w. h. p. all clients will have a fading coefficient to one of the cluster heads that is at least equal to $f_M$. 

In addition, we impose on the nodes the time division scheme of Section~\ref{subsection:time_division}: time is divided in frames, and each frame in $9$ slots. At any time during a slot, only a single node (either a cluster head or a client node) from each cell of the corresponding sub-lattice is allowed to transmit, and with maximum power. Since the receiver necessarily lies in the same cell, the lower bound on the SINR of Lemma~\ref{lemma8} continues to hold. Therefore, if the transmitter transmits with rate $f_R(\gamma_{\mathrm{min}}(n))$, where $\gamma_{\mathrm{min}}(n)$ is given by (\ref{eq14}), w. h. p. all transmissions will be successful. 

By (\ref{eq20}), there are less than $27 n^{1-d} \log n$ client nodes in each slot. We divide each slot in $2 \times [27 n^{1-d} \log n]$ time intervals, each of which is devoted to the transmission of a packet either from or to a client node. Each stream of data is guaranteed a rate of communication equal to $\lambda(n)= f_R(\gamma_{\mathrm{min}}(n))[2 \times 27 n^{1-d} \log n]^{-1}$. Multiplying by $2n$ for the total number of streams, and substituting for $\gamma_{\mathrm{min}}(n)$ from Lemma~\ref{lemma8}, we arrive at the lower bound (\ref{eq4b}).

\section{Hybrid Networks}
\label{section:hybrid}

Because of the similarities between cluster and hybrid networks, the wireless nodes can use for their communication the scheme that was used in Section~\ref{section:cluster} for proving the lower bound (\ref{eq4b}). In particular, wireless nodes do not transmit to each other, but rather transmit directly to an access point nearby. The packet is then transmitted through the infinite capacity network to an access point close to its destination, and is then transmitted one more time through the use of the wireless interface to the destination. All the analysis of Section~\ref{section:cluster} goes through, if we substitute client nodes with wireless nodes and cluster heads with access points. The only difference is that, because each packet must be transmitted twice, the aggregate throughput is half the throughput achieved in cluster networks. The bound (\ref{eq5}) follows. 

To derive (\ref{eq6}), we consider the opposite extreme. In particular, we note that the $n$ wireless nodes are free to ignore the infrastructure of the access points, and establish a communication scheme using only themselves. This uniform traffic case was the subject of \cite{gupta1}, and later \cite{toumpis9}. For reasons of completeness, in the Appendix we define such a network and prove that indeed it can achieve an aggregate throughput equal to the lower bound of (\ref{eq6}).

Regarding upper bounds on the capacity, although we provide no formal proof, it is intuitively clear that, in the case $d>\h$, the bound (\ref{eq5}) is tight, up to a poly-logarithmic factor. Indeed, the aggregate throughput of packets using the infrastructure, even for part of their transport, cannot exceed the upper bound (\ref{eq4a}), and the aggregate throughput of packets not using the infrastructure is much less, on the order of $n^\h$, by \cite{gupta1}. By a similar argument, the bound (\ref{eq6}) is tight, up to a poly-logarithmic factor, when $d<\h$.

\section{Conclusions}
\label{section:conclusions}

We study wireless networks with four different traffic pattern: asymmetric, multicast, cluster, and hybrid. The common aspect of these traffic patterns is their non-uniformity: in each of them some nodes are required to either send or collect much more traffic than other nodes This lack of uniformity places a strain on the network, through the formation of bottlenecks that have the potential to reduce the capacity. We present lower and upper bounds on the capacity that hold with probability going to unity as the number of nodes in the network goes to infinity. In the interest of brevity, we also present a number of conceptually straightforward upper bounds with only intuitive justification. Our work quantifies the inherent capabilities of wireless networks to handle various types of traffic pattern non-uniformities, and provides useful guidelines to protocol designers, for creating protocols that perform close to the capacity, without being overly complicated. 

Recently, a number of tight capacity bounds have appeared that are based on stochastic geometry tools, and in particular tools from percolation theory~\cite{franc1}. An open question is whether it is possible to sharpen or extend the results presented here using such tools. Combining traffic non-uniformities with results from stochastic geometry is a promising but challenging task and so is the subject of future work.

\appendix
\label{appendix1}

Here we present, for reasons of completeness, a proof of a lower bound of the capacity under uniform traffic similar to that of \cite{gupta1} mentioned in the Introduction. Our setting is similar, but not identical to the setting of \cite{gupta1}, and notably assumes fading. The proof is based on intermediate results of theorems that appeared in the main text. As a result, it is very short.

Let a \textbf{uniform network} consist of $n$ identical \textbf{wireless nodes} $X_1$, $X_2$,$\ldots$, $X_n$, placed randomly, uniformly and independently in the unit square $\{(x,y):  |x|,|y| \leq \h\}$. There are $n$ traffic streams and each wireless node is the source of a single stream, and the destination of a single stream. A node cannot be the source and destination of the \emph{same} stream. Apart from this restriction, all other combinations of sources and destinations are equally probable.
All streams have the same data rate $\lambda(n)~\mathrm{bps}$.

We define the \textbf{capacity $C(n)$ of the network} as the supremum of all rates $\lambda(n)$ that are uniformly achievable by all nodes, multiplied by their number $n$. The following theorem holds:
\begin{Theorem}
\emph{In uniform networks the capacity $C(n)$ is bounded w. h. p. as follows:
\begin{equation}
C(n) \geq \left[ \frac{3 \alpha -6}{3 \alpha -5}\right] \left[ \frac{W q f_m 5^{-\frac{\alpha}{2}}}{8600 \Gamma \log 2}\right]\frac{n^\frac{1}{2}}{(\log n)^\frac{3}{2}}.
\label{eq22}
\end{equation}
\label{theorem5}}
\end{Theorem}

\begin{Proof}{}
Let us divide the square region into a lattice of $g(n)=\frac{n}{18\log n}$ cells, as in Section~\ref{subsection:lattice}. The results \textbf{(i)}, \textbf{(ii)}, \textbf{(iii)} of that section continue to hold, with the understanding that the result \textbf{(iii)} applies to the communication of two wireless nodes using a third wireless node as relay. 

Furthermore, let us use the routing protocol of Section \ref{subsection:routing_protocol}, where now the destination node is actually another wireless node. To bound the number of traffic streams that each cell must support, let $M_j$ be the number of nodes with the same $x$-coordinate with cell $j$, and $N_j$ be the number of nodes with the same $y$-coordinate with cell $c_j$. Working as in Section~\ref{subsection:routing_protocol}, we readily have that, w. h. p., for all $j$, $M_j \leq \frac{9}{\sqrt{2}} \sqrt{n \log n}$, and by symmetry we also have $N_j \leq \frac{9}{\sqrt{2}} \sqrt{n \log n}$. Each node is the source and the destination of a single traffic stream, and therefore the number of routes supported by each cell is bounded w. h. p. as follows:
\begin{equation*}
\forall j,~~r_j\leq M_j+N_j \leq r_\mathrm{max}(n) \triangleq 9\sqrt{2} \sqrt{n \log n}.
\end{equation*}

If nodes use the time division scheme of Section~\ref{subsection:time_division}, then the bound (\ref{eq14}) on the SINR of each reception holds. Most of the routes going through a cell $c_j$ will require one reception, however a few of these, in particular those whose destination lies in cell $c_j$, may require three receptions.

Putting everything together, we have that each route, and its associated node, is guaranteed a rate of communication $\lambda(n)= f_R(\gamma_{\mathrm{min}(n)}) \left[ 3 \times 9 \times r_{\mathrm{max}}(n)\right]^{-1}$. Multiplying by $n$, and substituting for $r_{\mathrm{max}}(n)$ and $\gamma_{\mathrm{min}}(n)$, we see that our scheme achieves an aggregate throughput equal to the lower bound (\ref{eq22}). Since the capacity is the supremum of the aggregate throughputs of \emph{all} possible schemes, it will necessarily be greater than the aggregate throughput of our scheme, and the result follows.
\end{Proof}

\end{document}